\documentclass[doublecol,graphicx,times]{epl2} 
\usepackage{graphicx}
\usepackage{amsmath}
\usepackage{amsfonts}
\usepackage{amssymb}
\usepackage{amsbsy}

\def\sci{{Science}}

\def\asr{{Adv.\ Space Res.}}

\def\grl{{Geo\-phys.\ Res.\ Lett.}}

\def\jgr{{J.\ Geo\-phys. Res.}}

\def\prl{{Phys.\ Rev.\ Lett.}}

\def\pop{{Phys.\ Plasmas}}

\def\prl{{Phys. Rev. Lett.}}

\def\|{{\sss\parallel}}

\def\sss{\scriptscriptstyle}
\title{Auroral evidence for multiple reconnection in the magnetospheric tail plasma sheet}
\shorttitle{Multiple reconnection} 

\author{R. A. Treumann,\inst{1,2} C. H. Jaroschek\inst{3} \and R. Pottelette\inst{4}}
\shortauthor{R. A. Treumann, C. H. Jaroschek and R. Pottelette}

\institute{ 
  \inst{1} Department of Geophysics, Munich University, Theresienstr. 41, D-80333 Munich, Germany\\                   
  \inst{2} Department of Physics and Astronomy, Dartmouth College, Hanover, NH 03755\\
  \inst{3} Department of Earth and Planetary Sciences, Tokyo University, Tokyo, Japan\\
  \inst{4} CETP/CNRS St. Maur des Foss\'es, Cedex, France}
  

\abstract{We present auroral evidence for multiple and most probable small scale reconnection in the near Earth magnetospheric plasma sheet current layer during auroral activity. Hall currents as the source of upward and downward field-aligned currents require generation of the corresponding electron fluxes. The auroral spatial ordering in a multiple sequence of these fluxes requires the assumption of the existence of several -- and possibly -- even many tailward reconnection sites.  
}

\begin{document}

\maketitle
\noindent In the past three decades overwhelming evidence has been accumulated for reconnection in the tail of the magnetospheres of Earth and the other magnetised planets of the solar system to be the main energy release mechanism in magnetospheres. With very high probability reconnection is also in action in other astrophysical magnetised systems like the solar corona, where it may occur in solar flares, being a candidate of heating and accelerating the solar wind. If this turns out to be true, reconnection can be expected to participate in acceleration of stellar winds and to occur almost everywhere in interacting hot magnetised plasmas in the universe. Recently it has even been identified \cite{phan2007} in a thinning interplanetary current sheet in the magnetosheath, the transition region between Earth's bow shock wave and the outer boundary of Earth's magnetosphere, the magnetopause, where it has been made responsible for plasma heating \cite{retino2007}. This suggests that reconnection is a serious candidate even for the evolution of turbulence in a hot magnetised plasma and  is therefore also expected to occur in the transition from stellar winds to the interstellar media like the heliospheric heliosheath \cite{treumann2008}. 

In the magnetospheric tail the energy release by reconnection, which transforms the magnetic energy -- stored in the solar wind driven magnetospheric convection -- into plasma heating and injection into the inner magnetosphere, signs responsible for magnetospheric substorms and the occurrence of aurora. Still, even though reconnection has been identified in the tail (for a recent example see  \cite{xiao2007}) this is generally acknowledged, there is no consensus on whether reconnection in the tail just provides the energy for auroral processes (with other processes being responsible for aurorae), or whether the reconnection site directly feeds these processes. In the first case, the main auroral processes would result from other effects like tail current disruption, happening much closer to Earth, or simply from Alfv\'en waves that dissipate their energy in the upper ionosphere. In the second case, field aligned currents (or current pulses) flowing from the tail reconnection site into the ionosphere would directly build the upward-downward auroral current system. The presence of this upward/downward auroral field aligned current system has been inferred from ground-based and proved by {\it in situ} observations from low altitude spacecraft like VIKING, Freja \cite{marklund2001} and FAST \cite{carlson1998}. The presence of field aligned electron fluxes at the poleward plasma sheet boundary at $\sim$14 R$_E$ (geocentric Earth radii) distance in the tail has also been indirectly inferred long ago first from AMPTE IRM observations of locally excited electron plasma waves  \cite{schriver1990} for which electron beams of $\sim$keV energy sign responsible, suggesting a direct connection between the tailward reconnection site and the auroral region. However, the close-to-Earth observations  cannot distinguish between near and far Earth sources. The problem lies in the small-scale structure of the auroral current system and auroral phenomena as well as the lack of a viable tail reconnection model that explains how an auroral current system, which reproduces the  observations near Earth, is generated by reconnection in the thin more distant tail plasma sheet current layer when being fed by plasma inflow from the magnetospheric lobes. 
\begin{figure}[ht]
\centerline{{\includegraphics[width=8.6cm]{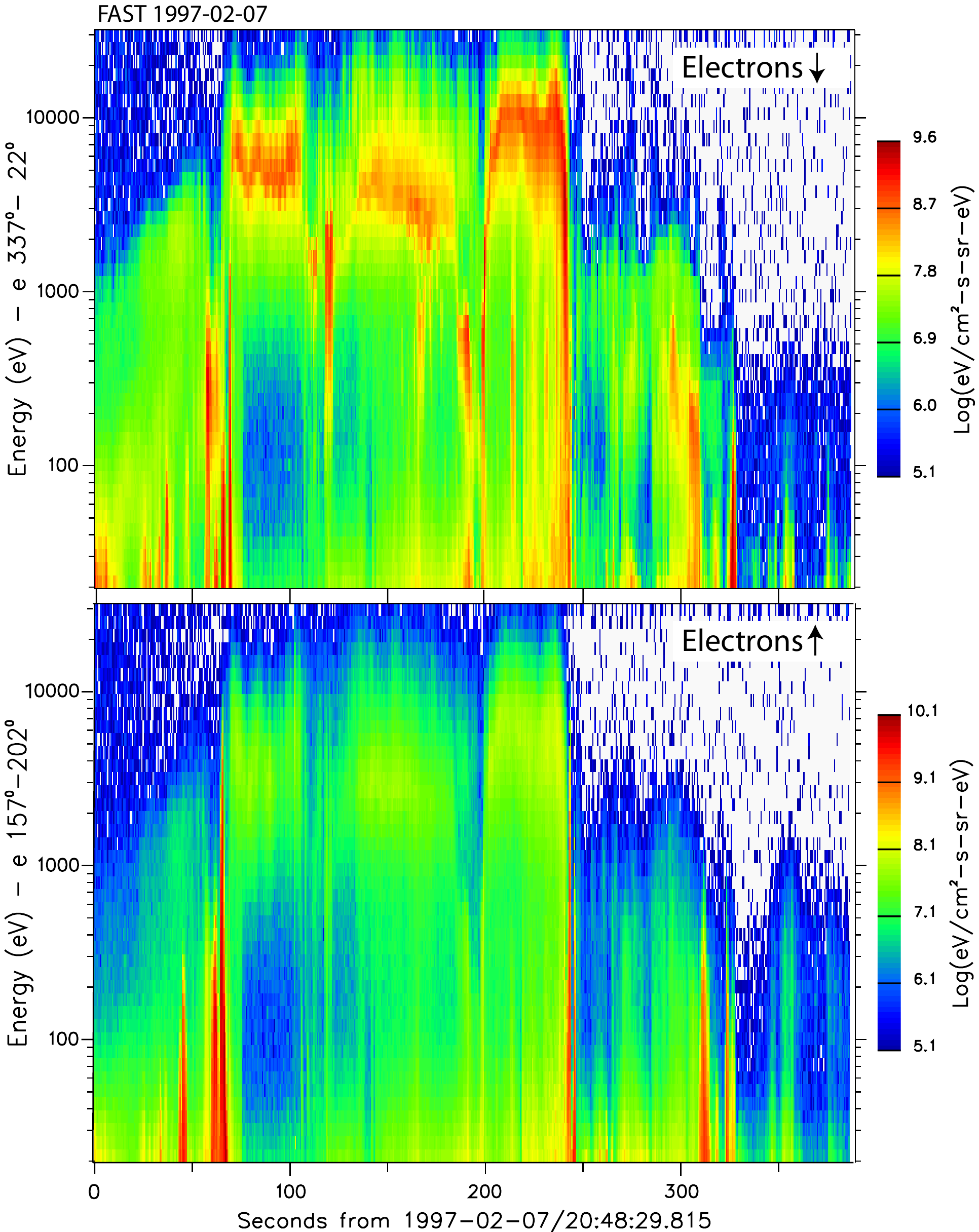}}}
\caption{Auroral electron energy fluxes measured by the FAST spacecraft at an altitude of roughly 4000 km above active aurorae in the high latitude terrestrial ionosphere. The figure shows a $\sim$6 min long spacecraft passage across the active auroral region from South to North on February 7, 1997. Given are the colour coded differential fluxes in eV/(cm$^2$s sr eV). Auroral activity started roughly 30 s before measuring time with main activity restricted to the time period as shown here.  The top panel shows the downward (parallel to the terrestrial magnetic dipole field in the northern hemisphere) electron flux. The lower panel shows the upward (anti-parallel) electron flux. Note the different colour codings on the right of the panels. Several broad regions of high energy downward fluxes are embedded into narrow regions of intense low energy upward electron fluxes. The image of the high energy downward fluxes in the bottom panel of the upgoing electrons are caused by electron backscattering of downward electrons in the auroral region and do not represent genuine upward accelerated auroral particles.}\label{aurorec-fig1}
\end{figure}

Figure \ref{aurorec-fig1} shows the typical example of a FAST spacecraft passage (at $\sim $4000 km altitude above Earth) through the magnetically connected to ground active auroral region. 
The two panels shown refer to downward (upper panel, density $N$$\sim$10$^4$m$^{-3}$, temperature $T$$\sim$10 keV) and upward (lower panel, $N$$\sim$10$^6$m$^{-3}$, temperature $T$$\sim$100 eV) electron fluxes, respectively, along the ambient geomagnetic dipole field. Downward (upward) current densities ${\bf j}_\|$ are of the order of $\sim$$10^{7-8}$A/m$^2$ ($\sim$$10^{6}$A/m$^2$). The upward current being distributed over a much wider latitude range. The passage lasted for $\sim$6 min, at a spacecraft orbital velocity of $\sim$5 km/s covering a distance of $\sim$1500 km or $\sim$6$^\circ$ in invariant latitude from South to North, almost the entire northern-hemispheric auroral region. 
Mapping it out into the magnetosphere, it corresponds to an equatorial distance of say $\sim$12-30 R$_E$ , a distance range that depends on the magnetospheric model used but spans quite a large range. 

The problem is buried in the fact that the entire event is by no means one single auroral event but consists of at least five and, presumably, even up to ten closely related events, each of them bounded by low energy upward electron fluxes which enclose the downward high energy electrons. Since the electrons must follow the magnetic field, each of the events is enclosed by magnetic field lines. Hence, when Figure \ref{aurorec-fig1} depicts a stationary upward and downward current system, then this system must be built of a sequence of separate auroral events which are ordered in a chain from South to North across the auroral zone. If, on the other hand, each of them is related to reconnection in the tail plasma sheet, then this reconnection on its own will consist of a series of reconnection zones that are located in the plasma sheet at increasing radial distance from Earth approximately covering the above estimated distance range of the entire auroral region. 
\begin{figure}[ht]
\centerline{{\includegraphics[width=8.5cm]{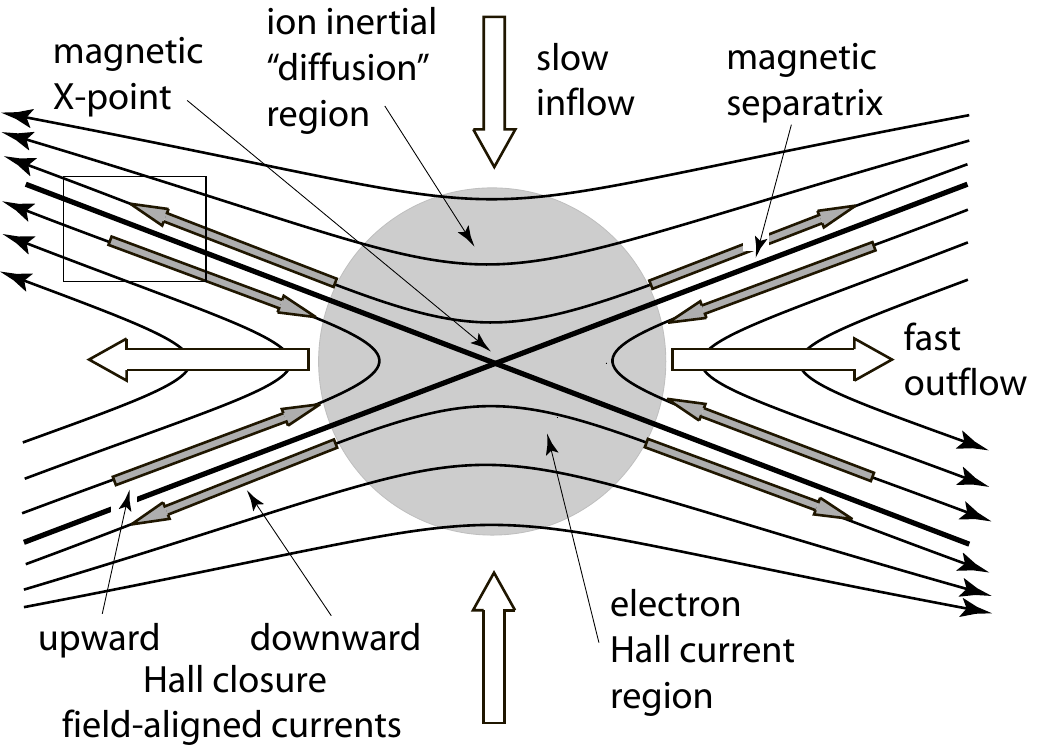}}}
\caption{Schematic of the reconnection site in a thin current layer of width  $\sim\lambda_i=c/\omega_{pi}$ (ion inertial length). Inflow is slow, outflow is fast. The ion "diffusion" region (grey) being of radius $\lambda_i$ contains unmagnetised ions. The field is transported by the cross-field drifting electrons, giving rise to Hall currents and Hall fields. Hall current closure is achieved  by field aligned currents along the magnetic field lines. The part accessible from the northern hemispheric auroral region is shown boxed.}\label{aurorec-fig2}
\end{figure}

It would, of course, also be possible that we were dealing here not with a stationary process but with a temporarily highly variable state which maps the unstable dynamics of one single reconnection zone in the tail. This possibility cannot completely be excluded. It is even not unreasonable as reconnection might in principle be a highly non-stationary state of the near-Earth plasma sheet, in particular because it is driven from the outside. In this case the spacecraft can be considered to be stationary, encountering the event from its southern edge and the event passing over it in southward direction, which is in agreement with the view that the reconnection in the tail ejects the plasma earthward, shortening the plasma sheet and dipolarising the magnetic field.  At the same time the reconnection site must oscillate back and forth with its downward current feet passing many times over the spacecraft in order to reproduce the event. 
\begin{figure}[ht]
\centerline{{\includegraphics[width=8.5cm]{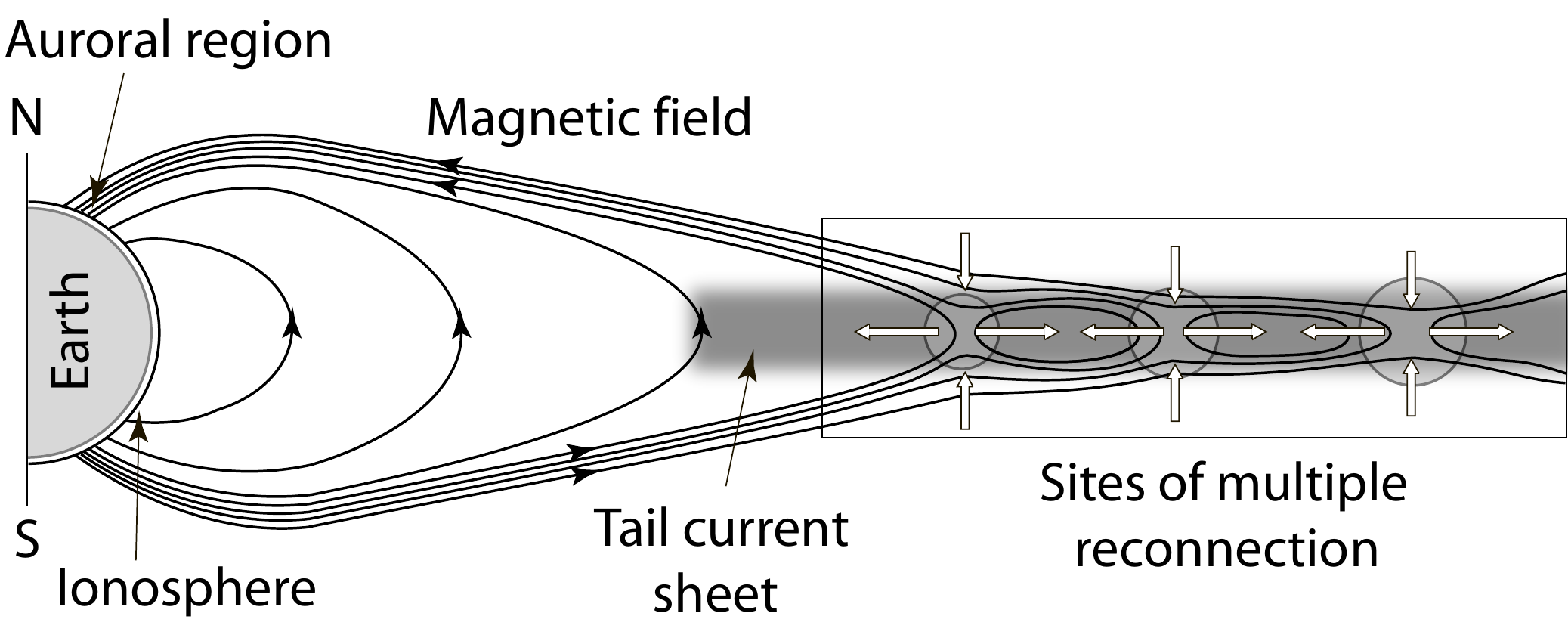}}}
\caption{Sketch of inner magnetospheric tail geometry (not to scale!) with tail current sheet and three adjacent reconnection sites. The three ion inertial diffusion regions are shows as circles. The presence of three reconnection sites (boxed here) requires that islands of closed magnetic fields form, known as `magnetic nulls' or 'plasmoids' between the \textsf{X} points. Such a chain of reconnection sites is typical for tearing modes. Here the difference is in the particular geometry that the earthward field lines are rooted in the ionosphere and body of Earth. This has consequences for the aurora. }\label{aurorec-fig3}
\end{figure}

Even though, at the current state of the discussion, this is a viable interpretation, we will take the former point of view and assume that the picture given in Figure \ref{aurorec-fig1} refers to a quasi-stationary state of several adjacent regions of auroral activity, which immediately raises the problem of how this can be possible in view of the traditional reconnection paradigm illustrated in Figure \ref{aurorec-fig2}. Here reconnection is basically two-dimensional, forms a magnetic \textsf{X}-point surrounded by the (circular) ion inertial region of radius $\lambda_i=c/\omega_{pi}\lesssim$1 R$_E$, being the site of some (unidentified) kind of ion diffusion and, since ions are inertia dominated and thus effectively unmagnetised, is also the site of Hall currents \cite{sonnerup1979,vaivads2004,treumann2006a,eastwood2007}. These are the consequence of the continuation of the inflow of magnetised electrons with velocity ${\bf V}_E={\bf E\times B}/B^2$ from North and South (not shown in the figure). The Hall electrons carry the magnetic field to which they are tied and escape from the ion inertial region (together with the magnetic field) to the left and to the right after reconnection took place. Closure of the Hall currents can be provided only by field-aligned currents flowing out and in as shown in the figure. The field-aligned currents correspond to electrons flowing in from and out to the environment connecting the tail reconnection site to the ionosphere.

Aside from the general problems in reconnection physics \cite{biskamp2000} and in particular Hall reconnection \cite{drake2008}, the problem of how to achieve a sufficiently thin current sheet, the dimensionality of reconnection (which in this model is assumed to be two) and a large number of further only partially solved complications, which require the use of numerical simulation techniques, the application to the auroral case encounters serious difficulties. The section of the reconnection site that maps down to the ionosphere is shown as the small box in the upper left corner. Accordingly, a single reconnection region should map in the ionosphere to just one pair of downward/upward field aligned currents. On the northern side of the auroral region the currents should flow into the ionosphere, while on the equatorward side they should flow out. Transforming to electrons, the northern side should exhibit upward electron flux, lifting the low energy ionospheric electron component up into the magnetosphere, while the heated electrons that have passed the inertial region will flow down into the southern part of the auroral region. Clearly this is incompatible with observations like those shown in Figure \ref{aurorec-fig1}. However, simulations of thin current sheets (for instance \cite{jaroschek2004} and others) show that in a sufficiently large box several reconnection regions evolve similar to the tearing mode. These reconnection sites form a chain of \textsf{X}-points and islands (nulls or plasmoids) which are in mutual motion and interact with each other.

Figure \ref{aurorec-fig3} exhibits a few interesting properties of multiple reconnection events in the geomagnetic tail. The first and simplest property is that the acceleration and ejection of plasma from each reconnection site into both directions to the right and left implies that the reconnection sites are not independent. Their interaction consist in the collision, retardation and mixing of the two plasma streams ejected into opposite directions from two adjacent reconnection sites. Strong reconnection in one place may in this way suppress weak reconnection in another place. Since two magnetised plasmas approach each other, we have a typical moving magnetic mirror configuration which, by the Fermi mechanism, is capable of accelerating particles. This is well known; it has recently been described in other places \cite{drake2003,drake2006,jaroschek2008}. A more efficient acceleration mechanism \cite{jaroschek2004} is by the reconnection electric fields if particles ejected from one reconnection site catch up with the reconnection electric field of another site and experience additional acceleration thereby providing a mechanism of producing high energy power law tails on the electron distribution function. These high energy electrons, when precipitating into the auroral ionosphere, cause the observed on Earth and Jupiter auroral X-ray emission by collisions with the neutral atmospheric constitutent \cite{gladstone2002,borg2007}. 
\begin{figure}[t]
\centerline{{\includegraphics[width=8.5cm]{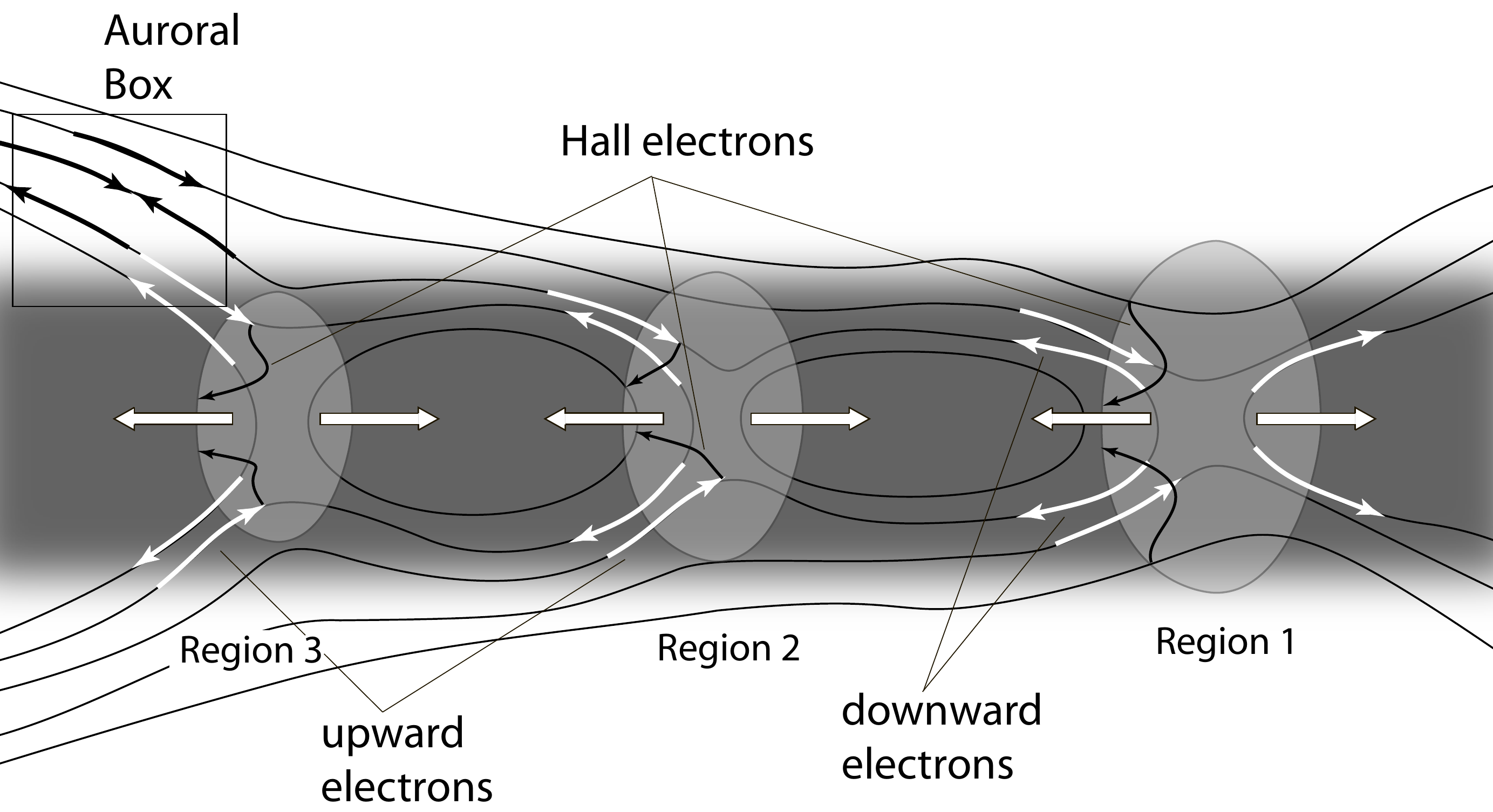}}}
\caption{Zoom of the three reconnection sites in Figure 3. The sites are numbered with decreasing distance from Earth. Straight white arrows are plasma outflow from reconnection sites (same as in Figure 3). The white arrows along the field lines show the direction of the upward and downward electron fluxes   which close the Hall currents (black arrows in ion diffusion regions are the corresponding Hall electron fluxes). The box in the upper left corner shows the electron fluxes that arrive in the auroral region.}\label{aurorec-fig4}
\end{figure}

Our main interest here is in the lower energy current carrying electrons. A short chain of reconnection sites consisting of three sites in the tail is sketched in Figure \ref{aurorec-fig4}. Region 1 is farthest away from Earth and has the largest extension because $\lambda_i$ increases with distance from Earth as the result of the radial decrease in plasma density. A few Hall electron flow lines (black arrows) in the reconnection regions are indicated in Figure \ref{aurorec-fig4}. These Hall electrons are fed by upward electron inflow along the magnetic field from the ionosphere drawn as white arrows from left and feed electrons into the aurora by downward flows (white arrows pointing to the left). Note that the upward fluxes from the ionosphere are located on the poleward field line of the corresponding reconnection site, while the downward fluxes are located on the equatorward field lines. Due to the particular geometry of the magnetospheric tail, the upward flux on the farthest northern field line connects to the outermost tailward reconnection site while the lowest latitude connection is to the innermost reconnection site that is located closest to Earth, and these fluxes are downward. 

However, what sequence of fluxes really arrives at and leaves from the auroral ionosphere is shown in the (northern) auroral box on the upper left. The region between the two outermost (northern and equatorward) auroral field lines contain mixed electron fluxes upward and downward depending on to which reconnection site the field line is connected. In particular, the field line which provides the upward flowing electrons to the Hall current in Region 3 also connects to Region 2 where it participates in reconnection and picks up those downward electrons that are leaving Region 2 in order to close the Hall current.  Similarly, the field line that provides upward electrons for the Hall current in Region 2 also connects to Region 1 where it reconnects and serves as guide for the downward accelerated Hall electrons from Region 1. One therefore expects that in the zone between the two outermost field lines upward and downward electron fluxes do not necessarily follow the naive sequence of Figure \ref{aurorec-fig2} but may mix. Observation of mixing of upward and downward auroral electron fluxes thus provides evidence for multiple reconnection taking place in the magnetospheric tail plasma sheet. The spacecraft is flowing from South to North. Hence, in a reconnection related reading the figure must be read from the right, i.e. beginning with Region 1, the outermost reconnection site and the northernmost field line. 
\begin{figure}[t]
\centerline{{\includegraphics[width=8.5cm]{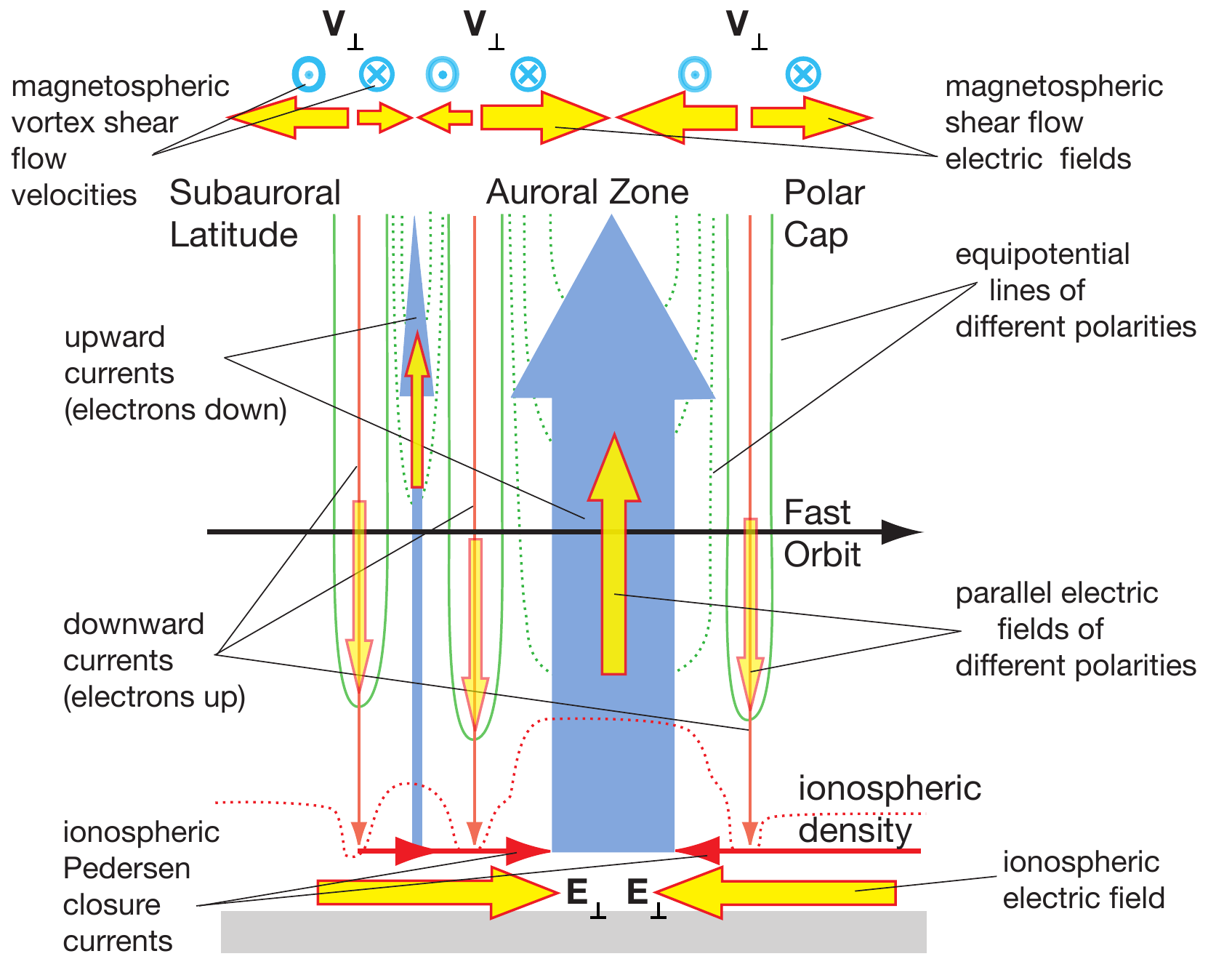}}}
\caption{The current system in the aurora inferred from the low altitude spacecraft when crossing a part of the auroral region as is believed to be related to the equatorial northern hemispheric multiple reconnection in the tail. Shown are the electric equipotentials which are assumed to be generated by topside convective shear flows  ${\bf V}_\perp$ \cite{carlson1998,singh2000} as indicated in the upper part of the figure. These are related to diverging or converging convection electric fields. At low altitudes below spacecraft path the potentials deviate from the magnetic field lines to close and produce field-aligned electric fields which accelerate electrons upward, causing downward currents, or downward, causing upward currents. The aurora is located in the upward current region caused by downward electrons. The currents close via the ionospheric Pedersen current parallel to the ionospheric electric field ${\bf E}_\perp$ at the bottom of the ionosphere.}\label{aurorec-fig5}
\end{figure}

Closer inspection of the sequence of electron fluxes in Figure \ref{aurorec-fig1} reveals the following:
Let us begin with the first large (from right) event at 245 s (skipping the few small po\-le\-ward events). It starts with a short intense upward electron burst (lower panel) that is equatorwards followed by intense downward electron fluxes at about 240 s in coincidence with further upward electrons.  Sufficiently intense upward electrons are present earlier from about $\sim$225 s which might partially be due to southward motion of the active aurora corresponding to the slow inward displacement of the reconnection region. These upward electrons coincide with several (3-4) bursts of downward electron injections of energy in the range of $\sim$100 eV (upper panel), which may be interpreted as a typical case of mixing of downward electrons from the main reconnection site and other close-by but further out reconnection regions. The downflowing (Hall) electrons at the source (reconnection site) have typical energies of a few 10-100 eV (being identified as reconnection electrons that do not belong to the above mentioned dilute high energy component that causes X ray emission).  Almost every downgoing event (upper panel), which is characterised by fluxes of $\sim$keV energy, starts with an increase in electron energy. This well known fact \cite{carlson1998} is interpreted as the entrance of downflowing electrons into a field-aligned quasi-stationary electrostatic accelerating potential (upward electric field) at auroral altitudes between 2000 km and 8000 km above ground the presumably cause of which are topside magnetospheric shear flows \cite{carlson1998,singh2000}. (Figure \ref{aurorec-fig5} shows a sketch of the ionospheric part of the electric field and current system deduced from the central part of the data in Figure \ref{aurorec-fig1}.)

Before 220 s the upward fluxes are week and the event is dominated by a single reconnection region. However, the structure of the entire event from 195-245 s is clearly complex being divided into 3-4 sub-events that are caused by the overlap of electrons going up and coming down. The injection of electrons at the equatorial end of this event at 195 s can be understood by the spacecraft briefly catching up the adjacent field line that already belongs to the next reconnection site. This field line is briefly lost and caught again (after five seconds at 190 s) further equatorward where it initiates the next event that is passed by the spacecraft. This is again a very complex event as seen from the upper panel, experiencing several injections of electrons from other reconnection sites. Its equatorward boundary somewhere around 130 s is  not marked by any spectacular signature in the electrons, in particular not by upward electron fluxes. 

Equatorward the next event (between 75 s and 110 s) is rather quiet and stable. Interestingly, it lacks upward electron fluxes at its northern boundary at 110 s. Its equatorward boundary is a sequence of downward electron bursts mixed with upward electron injections that partially overlap. Thus the whole event is rather complex. The absence of upward fluxes at the northern boundary and the bursts at the southern boundary cannot be brought into an orderly picture. From 0-60 s the latter mix into a broader equatorward region of lower energy downward dominated electron fluxes. The easiest explanation for this event is that it represents a complex probably three-dimensional reconnection structure.   

The above description is in relatively good agreement with the model of multiple tail reconnection displayed in Figure \ref{aurorec-fig4}. However, a number of caveats should be noted. The first concerns the assumed stationarity of the model. Reconnection, in particular the solar wind driven tail reconnection, is most probably a nonstationary process. It takes place under varying solar wind and magnetospheric convection conditions and storage of magnetic energy in the tail. Stationarity, as was assumed here, means that the system of reconnection sites in the tail remains intact for the time of the auroral event, in our case for $\sim$6 min. Even during this time, acceleration and interaction of the reconnection sites will cause displacements of the reconnection sites relative to each other, which is neglected in our simplified considerations. It may, however, also contribute to variations in the field aligned current and flow systems which is not considered here and complicates the picture.

In addition, the proposed multiple reconnection model is two-dimensional, which might be a serious oversimplification of the reconnection process. Various simulations suggested that reconnection is three-dimensional. It was found \cite{jaroschek2004} that the reconnection site was finite along the thin current sheet, being a few ion inertial lengths long. Similar results have been obtained in other simulations as well \cite{drake2008}. It is thus highly probable that the geometry of the upward and downward currents becomes complicated by the possibility of the field-aligned currents varying in the third dimension, which adds to the complexity of the auroral current structure that is caused in multiple reconnection. Three-dimensionality of reconnection enhances the probability of  dealing with multiple and possibly even multi-scale reconnection when observing the auroral plasma phenomena.

The main problems concern the dynamics of the Hall current system at the reconnection site, its closure, the generation of field aligned currents and, in particular, field aligned electron fluxes. Hall currents flow exclusively perpendicular to the magnetic field. Under normal conditions they are free of divergence forming vortices that close in themselves. In the ion inertial region they are forced to start at the convective electron entrance into and cease at electron leave from the ion inertial region. In order to avoid divergence they must close by non-Hall field-aligned currents \cite{sonnerup1979}. Since the upward current releases electrons from the reconnection site to the ionosphere, the downward currents on the poleward side need to provide the necessary electrons by sucking electrons up from the ionosphere. This can only be done by generating an electric field which accelerates ionospheric electrons the long distance up to the reconnection site. This field may be transported by a kinetic  Alfv\'en wave, but must first be generated in a process that is inherent to the reconnection mechanism. The Alfv\'en wave is kinetic because $\beta\sim 1$ in the ion inertial region and the transverse size is of the order of $\lambda_i$. At an average Alfv\'en speed of $10^3$ km/s, the travel time of this wave from the reconnection site at $\sim$15 R$_E$ to the ionosphere is roughly $\sim$100 s. This causes a delay between reconnection and the arrival of the upward accelerated ionospheric electrons at the reconnection site. The latter, being accelerated to $\sim$(0.1-1) keV, need only 2 s for travelling the same distance upward. The effect of this time delay on closure of the Hall currents is not known. It may retard the growth of reconnection, it may also cause some decorrelation between reconnection and the auroral response, increasing the complexity of the tail-aurora coupling.

Because of these caveats -- and also the uncertainties involved -- one cannot, at the current state of the art, expect complete agreement between the above model and observation. In particular, small-scale multiple reconnection in the magnetosphere is presumably genuinely three-dimensional. The model of the reconnection-aurora connection advocated in the present Letter is based on {\it in situ} auroral observations while being purely geometrical. Its numerical verification requires a mixture of global and local simulations with the local simulations being kinetic and allowing for the resolution of Hall current flow. Spatial three-dimensionality is not necessarily required as the Hall and field aligned currents can, in a simplistic model, be assumed to flow in the plane perpendicular to the magnetic field. Nevertheless, such simulations exceed current computing capabilities but may be expected to come into reach within the next decade.

{\small This research is part of a Visiting Scientist Programme at ISSI, Bern. CHJ has benefited from a JSPS Fellowship of the Japanese Society for the Promotion of Science. CHJ and RT thank M. Hoshino for hospitality, support and discussions. This research has also benefitted from a Gay-Lussac-Humboldt award of the French Government which is deeply acknowledged. Data access was through the French-Berkeley cooperative program on FAST supported by the French PNST programme. }

\end{document}